\documentclass[aps,prl, amsmath,preprint, preprintnumbers,superscriptaddress, floatfix, amssymb]{revtex4}

\usepackage{graphicx}
\usepackage{rotating}
\usepackage{dcolumn}
\usepackage{bm}
\usepackage{color}
\usepackage{mathptmx, textcomp}
\usepackage[latin1]{inputenc}
\usepackage{braket}

\begin{document}

\author{Elmar Haller}
\author{Russell Hart}
\author{Manfred J. Mark}
\author{Johann G. Danzl}
\author{Lukas Reichs\"ollner}
\author{Mattias Gustavsson}
\affiliation{Institut f\"ur Experimentalphysik and Zentrum f\"ur Quantenphysik, Universit\"at Innsbruck, Technikerstra{\ss}e 25, 6020 Innsbruck, Austria}
\author{Marcello Dalmonte}\affiliation{Institut f{\"u}r Theoretische Physik, Universit{\"a}t Innsbruck, Technikerstra{\ss}e 25, A--6020 Innsbruck, Austria}
\affiliation{Institut f\"ur Quantenoptik und Quanteninformation der \"{O}sterreichischen Akademie der Wissenschaften, Technikerstra{\ss}e 21a}
\affiliation{Dipartimento di Fisica dell'Universit\`a di Bologna, via Irnerio 46, 40127 Bologna, Italy}
\affiliation{Istituto Nazionale di Fisica Nucleare, via Irnerio 46, 40127 Bologna, Italy}
\author{Guido Pupillo}
\affiliation{ Institut f{\"u}r Theoretische Physik, Universit{\"a}t Innsbruck, Technikerstra{\ss}e 25, A--6020 Innsbruck,Austria}
\affiliation{Institut f{\"u}r Theoretische Physik, Universit{\"a}t Innsbruck, Technikerstra{\ss}e 25, A--6020 Innsbruck, Austria}
\author{Hanns-Christoph N\"agerl}
\affiliation{Institut f\"ur Experimentalphysik and Zentrum f\"ur Quantenphysik, Universit\"at Innsbruck, Technikerstra{\ss}e 25, 6020 Innsbruck, Austria}

\title{Pinning quantum phase transition for a Luttinger liquid of strongly interacting bosons}

\begin{abstract}
One of the most remarkable results of quantum mechanics is the fact that many-body quantum systems may exhibit phase transitions even at zero temperature\cite{Sachdev2001}. Quantum fluctuations, deeply rooted in Heisenberg's uncertainty principle, and not thermal fluctuations, drive the system from one phase to another. Typically, the relative strength of two competing terms in the system's Hamiltonian is changed across a finite critical value. A well-known example is the Mott-Hubbard quantum phase transition from a superfluid to an insulating phase\cite{Jaksch1998,Greiner2002}, which has been observed for weakly interacting bosonic atomic gases. However, for strongly interacting quantum systems confined to lower-dimensional geometry a novel type of quantum phase transition may be induced for which an arbitrarily weak perturbation to the Hamiltonian is sufficient to drive the transition\cite{Giamarchi2003,Tsvelik2004}. Here, for a one-dimensional (1D) quantum gas of bosonic caesium atoms with tunable interactions, we observe the commensurate-incommensurate quantum phase transition from a superfluid Luttinger liquid to a Mott-insulator\cite{Buechler2003,Pokrovsky1979}. For sufficiently strong interactions, the transition is induced by adding an arbitrarily weak optical lattice commensurate with the atomic granularity, which leads to immediate pinning of the atoms. We map out the phase diagram and find that our measurements in the strongly interacting regime agree well with a quantum field description based on the exactly solvable sine-Gordon model\cite{Coleman1975}. We trace the phase boundary all the way to the weakly interacting regime where we find good agreement with the predictions of the 1D Bose-Hubbard model. Our results open up the experimental study of quantum phase transitions, criticality, and transport phenomena beyond Hubbard-type models in the context of ultracold gases.
\end{abstract}

\maketitle

Ultracold atomic gases are a versatile tunable laboratory system for the investigation of complex many-body quantum phenomena\cite{Bloch2008}. The study of quantum phases and quantum phase transitions is greatly enriched by the possibility to independently control the kinetic energy and the interactions. In deep optical lattice potentials the many-body dynamics for a weakly interacting gas is, to a very good approximation, governed microscopically by a Hubbard Hamiltonian\cite{Jaksch1998} with a local onsite interaction energy $U$ and kinetic energy $J$, which corresponds to tunneling of atoms from one lattice site to the next. Experiments with Bose-Einstein condensates (BEC) of Rb atoms have demonstrated the quantum phase transition from a superfluid phase for large $J$ to an insulating Mott-Hubbard (MH) phase\cite{Greiner2002}. The transition between these two phases was obtained by quenching $J$ in a lattice of finite depth. Recent experiments with fermionic atoms have demonstrated the presence of a fermionic MH insulating state\cite{Joerdens2008,Schneider2008}, potentially opening the way to the study of high-temperature superconductivity in proximity of the MH phase in 2D.

\begin{figure}[t]
    \centering
    \includegraphics[width=12cm]{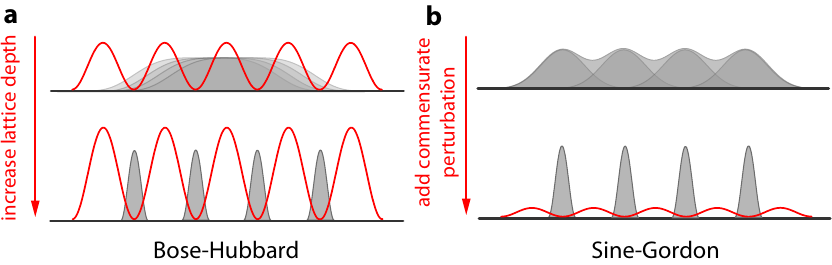}
    \caption{{\bf Comparing two types of superfluid-to-Mott-insulator phase transitions in 1D.} Schematic density distributions (grey) in the presence of a periodic potential (red solid line). {\textbf a}, Mott-Hubbard type quantum phase transition for weak interactions\cite{Greiner2002}. The system is still superfluid at finite lattice depth (top). The transition to the insulating state is induced by raising the lattice depth above a finite critical value (bottom). {\textbf b}, Sine-Gordon type quantum phase transition for strong interactions\cite{Buechler2003}. In the absence of any perturbation, the system is a strongly correlated superfluid (top). For sufficiently strong interactions, not necessarily infinitely strong, an arbitrarily weak perturbation by a lattice potential commensurate with the system's granularity induces the transition to the insulating Mott state (bottom).
\label{fig1}}
\end{figure}

While the focus in the study of quantum phase transitions in the context of ultracold atoms has so far been on Hubbard-type physics in the weakly interacting regime, novel quantum phenomena occur in lower dimensions, where the effects of quantum fluctuations and correlations are enhanced. In a 1D bosonic gas, strong repulsive interactions lead to the formation of a Tonks-Girardeau (TG) gas, where bosons minimize their interaction energy by avoiding spatial overlap and acquire fermionic properties\cite{Girardeau1960,Kinoshita2004,Paredes2004,Haller2009}. The addition of an arbitrarily weak lattice potential commensurate with the atomic density, i.e. $n \sim 2/\lambda$, where $n$ is the linear 1D density and $\lambda/2$ is the lattice periodicity, is expected to lead to a novel kind of quantum phase transition\cite{Giamarchi2003,Buechler2003}: the strongly correlated 1D gas is immediately pinned by the lattice and the superfluid TG phase is turned into an insulating, gapped phase. Figure~\ref{fig1} contrasts the Hubbard-type superfluid-to-Mott-insulator transition to this pinning transition. Given the universality of 1D quantum physics, the pinning transition will occur for interacting bosons as well as for fermions in 1D and has been discussed with respect to a variety of quantum models in low dimensions\cite{Giamarchi2003}.

The pinning transition is described by the (1+1) quantum sine-Gordon (sG) model, which is an exactly solvable quantum field theory, extensively studied in high energy, condensed matter, and mathematical physics\cite{Tsvelik2004}. The sG Hamiltonian reads
\begin{equation}
\mathcal{H}=\frac{\hbar v_s}{2 \pi}\int dx [(\partial_x \theta)^2 + (\partial_x \phi)^2 + \mathcal{V} \cos(\sqrt{4 K} \theta)]. \label{sG}
\end{equation}
\noindent Here, $\partial_x \theta$ and $\partial_x \phi$ are the fluctuations of the long-wavelength density and phase fields $\theta$ and $\phi$, respectively, of the hydrodynamic description of the 1D liquid with commutation relation $[\partial_x  \theta (x),\phi(y)]=i \pi \delta(x-y)$,  $v_\text{s}$ is the velocity of the soundlike excitations of the 1D gas, $\mathcal{V}=V n \pi/(\hbar v_\text{s})$ is proportional to the depth $V$ of a weak lattice\cite{Giamarchi2003,Buechler2003}, and $\hbar$ is Planck's
constant $h$ divided by $2 \pi$. For vanishing lattice $\mathcal{V}=0$, Eq.~\eqref{sG} describes a Luttinger liquid, where the strength of interactions is parameterized by the dimensionless parameter $K=\hbar \pi n/(m v_\text{s})$, which determines the long-distance power-law decay of the correlation functions, e.g. $\langle n(x)n(x')\rangle\sim 1/|x-x'|^{2K}$, with $m$ the atomic mass. The sG model with a weak but finite lattice predicts a quantum phase transition of the Berezinskii-Kosterlitz-Thouless (BKT) type between a superfluid state for $K>K_\text{c}=2$, where the shallow lattice is an irrelevant perturbation, to an insulating Mott phase for $K<K_\text{c}$, for which the spectrum is gapped for any value of $\mathcal{V}$.

While in general $K$ is a phenomenological parameter, in the case of a 1D bosonic gas it can be microscopically related to the Lieb-Liniger parameter $\gamma= m g /(\hbar^2 n)$, which characterizes interactions in a homogenous 1D system\cite{Lieb1963} (see Methods). Here, $g\simeq 2 \hbar \omega_{\perp} a_\text{3D}$ is the coupling constant of the 1D $\delta$-function interaction potential $U(x)=g \delta(x)$, where $\omega_\perp$ is the frequency of transverse confinement and $a_\text{3D}$ is the 3D scattering length. The strength of interactions, and thus $K$, can be tuned by varying $a_\text{3D}$ near a Feshbach resonance\cite{Chin2008}. The TG regime corresponds to $\gamma \gg 1 $. Using the relation between $K$ and $\gamma$, B\"uchler and coworkers\cite{Buechler2003} have shown that particles are pinned for experimentally accessible values of $\gamma > \gamma_\mathrm{c} \simeq 3.5$ in the limit of a vanishingly weak lattice. The pinning transition is expected to continuously transform into the MH-type quantum phase transition, which occurs for the weakly interacting gas when the lattice depth becomes sufficiently large. Here, using a quantum gas of caesium (Cs) atoms with tunable interactions confined to an array of independent 1D tubes (see Methods), we drive the superfluid-to-Mott-insulator phase transition by varying $\gamma$ and determine the phase boundary all the way from the strongly to the weakly interacting regime using modulation spectroscopy and measurement of transport. For shallow lattices under conditions of commensurability, we observe immediate pinning of the particles for strong interactions when $\gamma>\gamma_\text{c}$.

\begin{figure}
\begin{center}
 \includegraphics[width=12.0cm] {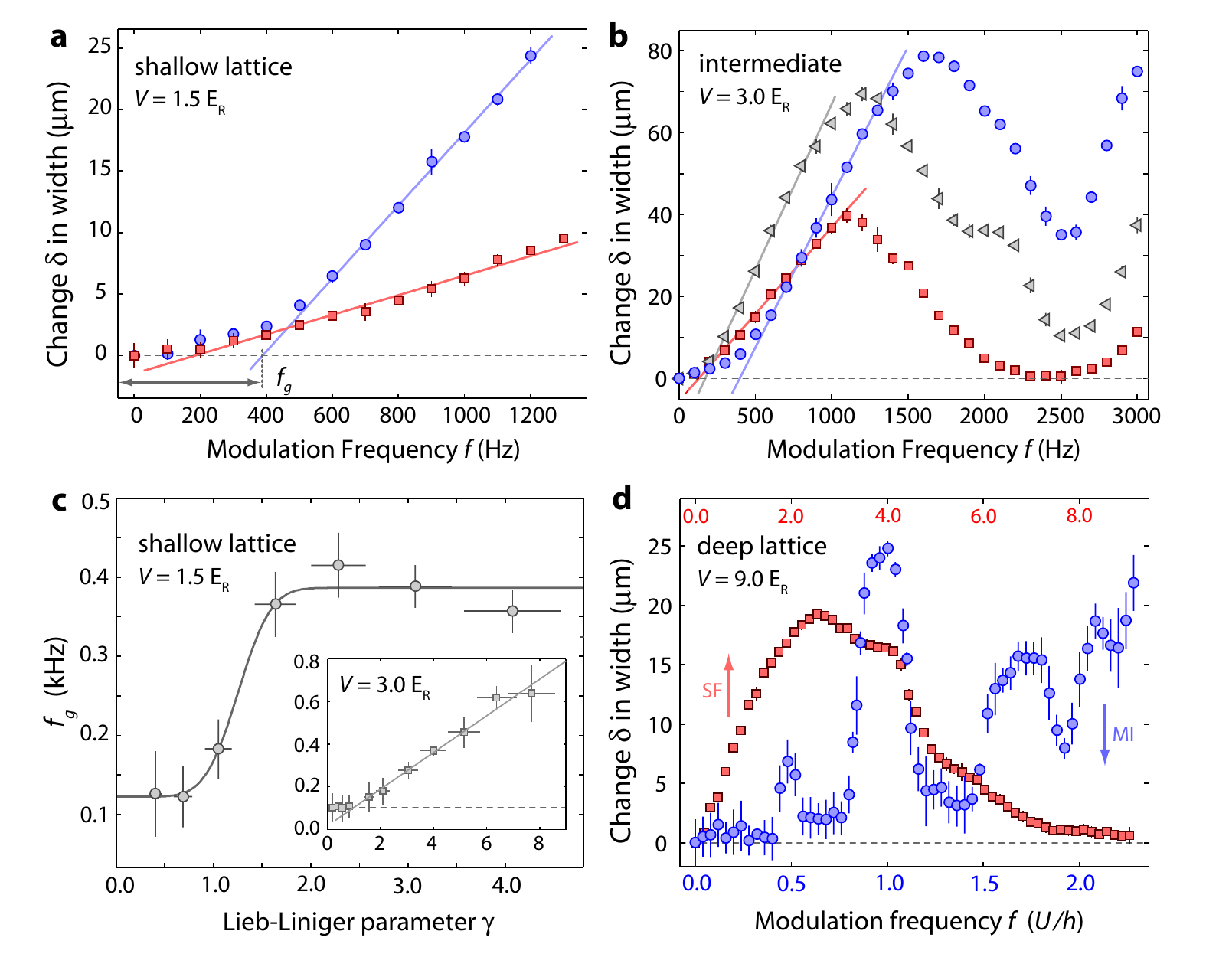}
 \end{center}
 \caption{\textbf{Modulation spectroscopy on bosons in 1D.} \textbf{a, b, d}, Excitation spectra for low, intermediate, and high lattice depth $V$. The change $\delta$ of the spatial width after amplitude modulation is plotted as a function of the modulation frequency $f$ for different values of $\gamma$. {\textbf a}, Characteristic spectra for $V=1.5(1) E_\text{R}$, one in the superfluid regime (squares, $a_\text{3D}=115(2) \ a_0$, $\gamma=1.0(1)$) and one in the Mott regime (circles, $a_\text{3D}=261(2) \ a_0$, $\gamma=3.1(2)$). The solid lines are linear fits to the high-frequency part of the spectrum. We determine the axis intercept $f_\text{g}$ as indicated. {\textbf b}, Spectra for $V=3.0(2) E_\text{R}$. The system is superfluid at $\gamma=0.51(6)$ (squares), while it exhibits a gap for $\gamma=1.6(1)$ (triangles) and $\gamma=4.1(3)$ (circles). Additional structure appears at higher frequencies due to the presence of band structure. {\textbf c}, Determination of the transition point for the case of the shallow lattice with  $V=1.5(1) E_\text{R}$. The frequency $f_\text{g}$ is plotted as a function of $\gamma$. The transition point is identified with the abrupt step. The solid line is an error-function fit to the data from which we determine the critical value $\gamma_{\text{c},V}$. The inset plots $f_\text{g}$ as a function of $\gamma$ for $V=3.0(2) E_\text{R}$. We identify the transition with the onset of the linear rise in $f_\text{g}$. {\textbf d}, Spectra for $V=9.0(5) E_\text{R}$. Here, we give $f$ in units of $U$. For comparatively weak interactions (squares, $\gamma=0.10(3)$) the spectrum exhibits a broad distribution characteristic of a superfluid (SF). For stronger interactions (circles, $\gamma=8.1(4)$) pronounced peaks develop, indicating the presence of a Mott insulator (MI). In \textbf{a, b, d}, the error bars for $\delta$ reflect the $1\sigma$ statistical error. In \textbf{c}, the error bars for $f_\text{g}$ are derived from the $1\sigma$ error on the fit parameters. The error for $\gamma$ results from the $1\sigma$ statistical error of the independent input variables and the spread of $\gamma$ due to the distribution of tubes.
 \label{fig2}}
\end{figure}

We first discuss our experiments in the strongly interacting regime. We start with a 3D Bose-Einstein condensate (BEC) of typically $1.3\times10^5$ Cs atoms without detectable thermal fraction in a crossed-beam dipole trap with magnetic levitation\cite{Kraemer2004} and initialize our system by creating a conventional 3D MH-state in a deep 3D lattice at $U/(6J) \approx 75$ with precisely one atom per lattice site\cite{Greiner2002}. The array of 1D tubes is obtained by reducing the lattice depth $V$ along one direction. Our procedure ensures that a majority of tubes has a near-commensurate number density (see Methods). A Feshbach resonance allows us to control $a_{\text{3D}}$ with a precision of $3$ $a_0$ limited by the presence of the magnetic field gradient. Here, $a_0$ is Bohr's radius. For the case of the shallow lattice, we probe the state of the system by amplitude modulation spectroscopy\cite{Stoeferle2004}. We determine the presence of an excitation gap $E_\text{g}$ by testing whether energy can be deposited into the 1D system at a given excitation frequency $f$. The lattice depth $V$ is modulated at $f$ by $25\%$ to $45\%$ for $40-60$ ms. After ramping down the lattice beams adiabatically with respect to the lattice band structure and after a levitated expansion time of $40-60$ ms\cite{Kraemer2004}, we detect the atoms by time-of-flight absorption imaging. We determine the spatial width of the atomic sample from a gaussian fit to the absorption profile and obtain the change $\delta$ of the spatial width compared to the unmodulated case as a function of $f$. Two typical measurements are shown in Fig.~\ref{fig2}(a), one in the superfluid phase and one deep in the 1D Mott phase at the same value for the lattice depth, $V = 1.5(1) E_\text{R}$, where $E_R=h^{2}/(2 m \lambda^{2})$ is the photon recoil energy. For weak interactions the system exhibits a linear increase for $\delta$ as a function of $f$, which we attribute to the superfluid character of the gas. For strong interactions, the increase, after a slow rise, shows a clear kink. We attribute the initial slow rise to excitation of residual superfluid portions of our inhomogeneous system and the sudden change in slope to the presence of an excitation gap. We associate the axis intercept $f_\text{g}$ obtained from a linear fit to the steep part of the spectrum with the frequency of the gap. To determine the phase transition from the 1D Mott state to the superfluid state, we repeat this measurement for a given depth $V$ as we scan $\gamma$ by changing $a_{\text{3D}}$. A typical result is shown in Fig.~\ref{fig2}(c). The gap closes as $\gamma$ is reduced. For values $V\le 2.0 E_\text{R}$, we determine the critical value $\gamma_{\text{c},V}$ at which the transition happens by an error-function fit to the data. Note that we always observe some small residual value for $f_\text{g}$ of about 120 Hz for weak interactions.

For comparison, we present in Fig.~\ref{fig2}(b) and (d) excitation spectra for an intermediate value of the lattice depth and for the case of a deep lattice, respectively. For $V = 3.0(2) E_\text{R}$ the spectrum shows additional structure for high frequencies as band structure comes into play. We find that for $V > 2.0 E_\text{R}$ the gap opens up approximately linearly as a function of $\gamma$ beyond a critical $\gamma_{\text{c},V}$, see inset to Fig.~\ref{fig2}(c). For deep lattices we recover the discrete excitation spectrum of the Mott phase in the Hubbard regime\cite{Greiner2002,Stoeferle2004} with a pronounced peak at $f=1.0 \ U/h$. Additional peaks\cite{Jaksch2005} can be found at $f=0.5 \ U/h$ and above $f=1.5 \ U/h$.

\begin{figure}[t]
\begin{center}
\includegraphics[width=12.0cm]{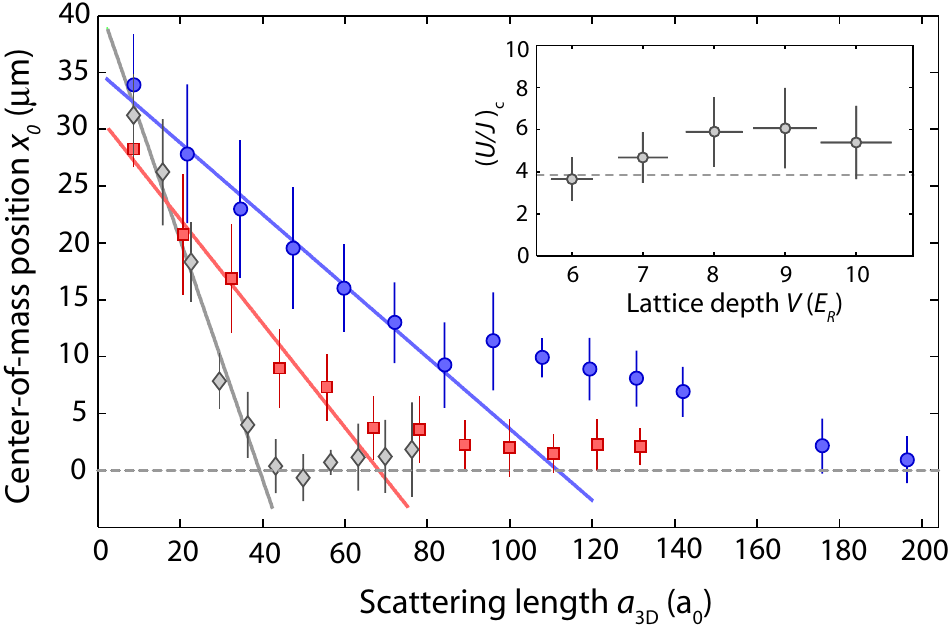}
\end{center}
\caption{\textbf{Transport measurements on the 1D Bose gas.} Center-of-mass displacement $x_0$ as a function of $a_\text{3D}$ for different values of $V$ ($V=9.0(5) E_\text{R}$ (diamonds), $V=5.0(3) E_\text{R}$ (squares), $V=2.0(1) E_\text{R}$ (circles)). We extrapolate the linear slope at small values for $a_\text{3D}$ and associate the transition point with the axis intercept. For the data with $V=2.0(1) E_\text{R}$ transport is not fully quenched as the condition of commensurability is not fulfilled for all atoms. All errors are the $1\sigma$ statistical error. The inset plots the mesured critical ratio $(U/J)_\text{c}$ at the transition point as a function of lattice depth $V$. The dashed line indicates the theoretical result $(U/J)_\text{c} \approx 3.85$ for the 1D Bose-Hubbard regime\cite{Rapsch1999}.
\label{fig3}}
\end{figure}

For the case of a deep lattice, we find that the state of the system is very sensitively probed by transport measurements\cite{Fertig2005,Mun2007}. A characteristic property of the Mott state is the inhibition of particle motion. In our experiment with the capability to tune interactions we expect the phase transition to manifests itself, at fixed $V$, through a strong suppression of transport when the strength of the interaction is raised above a certain critical value. Essentially, we test whether momentum can be imparted to the 1D system as a function of interaction strength. For a given $V$ we apply a weak axial magnetic force for a brief time to the interacting system, chosen such that the imparted momentum would be approximately $0.2 \hbar k$ if the system were non-interacting. Then, as a function of $a_\text{3D}$, we determine the center-of-mass displacement $x_0$ of the sample after a fixed time of flight. Fig.~\ref{fig3} shows that $x_0$ decreases monotonically with $a_\text{3D}$. For the case of a deep lattice with $V = 9.0(5) E_\text{R}$ the quenching of transport is abrupt. At a certain critical value for $a_\text{3D}$ transport is fully inhibited\cite{Altman2005,Guido}. We find the critical $a_\text{3D}$ by a linear fit to the decreasing data and by determining the axis intercept and derive from this a critical $\gamma_{\text{c},V}$ . Reducing the lattice depth to $V = 5.0(3) E_\text{R}$ and $V = 2.0(1) E_\text{R}$ leads to a less abrupt quenching of transport. For stronger interactions, the decrease starts to level off. Nevertheless, the initial decrease is still linear, allowing us to determine the critical $\gamma_{\text{c},V}$ by an extrapolation of the initially linear decrease to zero. The inset to Fig.~\ref{fig3} shows the measured critical ratio $(U/J)_\text{c}$ determined by our transport method as a function of lattice depth $V$. When we compare our results with the predicted value\cite{Rapsch1999} of $(U/J)_\text{c} \approx 3.85$ for the transition in 1D, we find a slight systematic overestimation of the transition point. This, however, is expected in view of e.g. the spatial inhomogeneity of the sample and the BKT-type nature of the transition in a finite size system.

\begin{figure}[t]
\begin{center}
 \includegraphics[width=14.0cm] {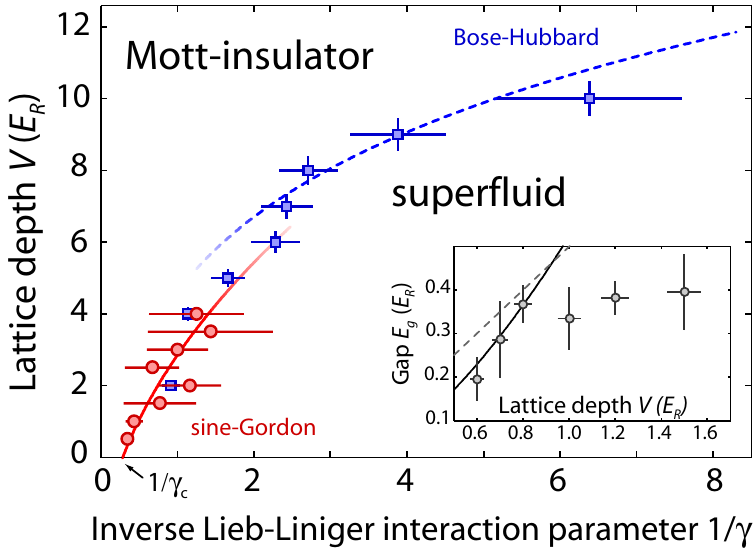}
 \end{center}
 \caption{{\bf Phase diagram for the strongly interacting 1D Bose gas.} Superfluid and Mott insulating phases in 1D versus inverse Lieb-Lininger interaction parameter $1/\gamma$ and optical lattice depth $V$ in units of the photon recoil energy $E_\text{R}$. The critical interaction parameter is $\gamma_\text{c}$. The data points (squares and circles) indicate the measured phase boundary. For strong interactions and shallow lattices we determine the transition by amplitude modulation spectroscopy (circles). Our measurements are in good agreement with the theory based on the sine-Gordon model (solid line, see Methods). For $\gamma > \gamma_\text{c}$ an arbitrarily weak lattice induces the transition to the insulating state. For weak interactions and deep lattices we probe the phase boundary by transport measurements (squares). Here, our measurements agree well with the prediction from the 1D Bose-Hubbard model (dashed line). For the modulation measurements, the error in $\gamma$ is derived from the $1\sigma$ error of the fit parameters. For the transport measurements, the error in $\gamma$ results from the $1\sigma$ statistical error of the independent input variables and the spread of $\gamma$ due to the distribution of tubes. The inset plots the measured gap energy $E_\text{g}= h f_\text{g}$ as a function of $V$ for $\gamma=11(1)$ and compares our data to the analytical result for finite $\gamma$ as given by the sine-Gordon model (solid line, see Methods). Also shown is the universal behavior $E_\text{g}=V/2$, which is valid for non-interacting fermions (dashed line).
\label{fig4}}
\end{figure}

We summarize our results in Fig.~\ref{fig4}, where we present the phase diagram  as a function of $1/\gamma$ and $V$. The set $\{ \gamma_{\text{c},V} \}$ defines the phase boundary between the 1D Mott insulator and the 1D superfluid. The measurements based on modulation spectroscopy cover a range from $V=4 E_\text{R}$ down to $0.5 E_\text{R}$ (circles), while the transport measurements extend from $V=2 E_\text{R}$ to $10 E_\text{R}$ (squares). In the weakly interacting regime, $1/\gamma > 2$, our data are in good agreement with the prediction of the MH model (dashed line). In the strongly interacting regime, $1/\gamma < 1$, the measured phase boundary extrapolates to a finite critical value $1/\gamma_\text{c}$ for the Lieb-Liniger parameter as the lattice depth $V$ is reduced to zero. Our results are in excellent quantitative agreement with the theory for a commensurate system based on the sine-Gordon model (solid line, see Methods), for which $\gamma_\text{c}=3.5$. We also find good agreement between our two types of measurement techniques in the intermediate regime ($V=2 E_\text{R}$ to $4 E_\text{R}$). Our results demonstrate the striking consequence of strong interactions in 1D geometry in the presence of a lattice: Beyond a critical value $\gamma_\text{c}$, an insulating Mott state exists for vanishingly small lattice depth $V$. The particles are immediately pinned by the lattice.

We measure a finite gap energy $E_\text{g}$ for $\gamma > \gamma_\text{c}$ in the regime of a shallow lattice. In the limit of $\gamma \to \infty$ and $V \to 0$ one would expect the simple relation $E_\text{g}=V/2$ as the bosonic system has become fully fermionized and the lattice effectively induces a band insulator of fermions\cite{Buechler2003}. In the inset to Fig.~\ref{fig4} we plot the measured $E_\text{g}$ as a function of $V$ at fixed $\gamma=11(1)$. For $V < 1 E_\text{R}$ our data is in good agreement with the analytical result for the gap energy at finite $\gamma$ (see Methods). Note that, for $V \geq 1 E_\text{R}$, we observe a deviation for $E_\text{g}$ away from the predicted values. This deviation occurs at rather shallow lattices. However, one does expect the curve to have a reduced slope for deeper lattices, for which $E_\text{g}$ becomes of order $U$ and is only weakly dependent on $V$.

Our results are a benchmark realization of quantum field theory models with tunable parameters in cold atomic systems. These results open up the experimental study of the out-of-equilibrium properties of sine-Gordon-type models. In particular, thermalization in integrable models beyond the Luttinger liquid model, quenches across quantum phase transitions, and their relations to the breakdown of the adiabatic theorem in low dimensions can now be investigated with full tunability of system parameters.

\subsection{Acknowledgements}

We thank W. Zwerger for discussions. We are indebted to R. Grimm for generous support. We gratefully acknowledge funding by the Austrian Ministry of Science and Research (Bundesministerium f\"ur Wissenschaft und Forschung, BMWF) and the Austrian Science Fund (Fonds zur F\"orderung der wissenschaftlichen Forschung, FWF) in form of a START prize grant and by the European Union through the STREP FP7-ICT-2007-C project NAME-QUAM (Nanodesigning of  Atomic and MolEcular QUAntum Matter) and within the framework of the EuroQUASAR collective research project QuDeGPM. R.H. is supported by a Marie Curie International Incoming Fellowship within the 7th European Community Framework Programme.

\section{Methods}\subsection{1D Bose gas in a weak optical lattice.}

In the absence of the optical lattice, $V=0$, the Luttinger liquid parameter $K$ can be expressed in terms of the Lieb-Liniger parameter $\gamma=g m /(\hbar^2 n)$ for all strengths of interactions\cite{Lieb1963,Cazalilla2004}. For $\gamma \leq 10$ and $\gamma \gg 10$ one gets $K\simeq \pi/\sqrt{\gamma-\gamma^{3/2}/(2\pi)}$ and $K\simeq (1+2/\gamma)^2$, respectively. The addition of a weak but finite commensurate optical lattice with $V \leq 1 E_\text{R}$ realizes the effective sine-Gordon Hamiltonian Eq.~\eqref{sG}. Using a perturbative renormalization group approach, the BKT transition line between the superfluid and the Mott-insulating phases can be derived in terms of $V$ and $\gamma=\gamma_{\text{c},V}$ as
\[
\frac{V}{E_\text{R}}=2\left(\frac{\pi}{\sqrt{\gamma-\gamma^{3/2}/(2\pi)}}-2\right).
\]
For small lattice depths, the integrable structure of the sine-Gordon model\cite{Zamolodchikov1979,Zamolodchikov1995} allows one to derive the following analytical expression for the dependence of the spectral gap $E_\text{g}$ on $V$ and $K$
\[
\frac{E_\text{g}}{E_\text{R}}=\frac{8\Gamma[\frac{\pi K}{2(2-K)}]}{\sqrt{\pi}\Gamma[\frac{1}{2}\frac{2+K(\pi-1)}{2-K}]}\left[\left(\frac{K^2V}{16E_\text{R}}\right)
\frac{\Gamma[1-\frac{K}{2}]}{\Gamma[1+\frac{K}{2}]}\right]^{\frac{1}{2-K}}.\]
Here, $\Gamma$ is the gamma function. For strong interactions $K \simeq 1$, the dependence of the gap on $V$ is linear, and $E_\text{g}$ approaches the free fermion value $E_\text{g} = V/2$. In the vicinity of $K = 2$, the gap closes exponentially approaching the BKT transition line.

\subsection{Deep lattice: the Bose-Hubbard model.}
In the weakly interacting regime $\gamma \ll 1$, for $V \gg 1 E_\text{R}$, when all atoms occupy the lowest vibrational state in each potential well of the lattice, the system can be described by the following Bose-Hubbard model\cite{Jaksch1998}
\[
H=-J\sum_{i}(b_{i}^{\dagger}b_{i+1}+h.c.)+\frac{U}{2}\sum_{i}b_{i}^{\dagger}b_{i}^{\dagger}b_{i}b_{i}.\]
Here, $b_i$ ($b_i^\dagger$) is the operator destroying (creating) a bosonic particle at the position of the $i^{\rm th}$-well, $J=4E_\text{R}(V/E_\text{R})^{\frac{3}{4}}\exp[-2\sqrt{V/E_\text{R}}]/\sqrt{\pi}$
is the hopping energy, and $U=\sqrt{2\pi}g (V/E_\text{R})^{1/4}/\lambda$ is onsite interaction energy. The quantum phase transition between a superfluid and a MH state occurs at\cite{Rapsch1999} $(U/J)_\text{c} \approx 3.85$, which determines a transition line in the ($V,\gamma$) - plane via \[
\frac{4V}{E_\text{R}}=\ln^{2}\left[\frac{2 \sqrt{2} \pi }{\gamma} \left( \frac{U}{J} \right)_\text{c}\sqrt{\frac{V}{E_\text{R}}}\right].\]

\subsection{Lattice loading and array of 1D tubes.}
We create a 3D optical lattice by interference of 3 pairs of counterpropagating dipole trap laser beams at wavelength $\lambda=1064.5$ nm with $1/e^2$ beam waists of $\sim 350 \ \mu$m. The atomic BEC, initially trapped in a crossed-beam dipole trap, is adiabatically transferred to the 3D lattice by exponentially ramping up the power in the lattice laser beams within $300$ ms. We create a 3D Hubbard-type Mott insulator with precisely one atom per site in the central region of the trap by adjusting the external dipole trap confinement. The array of vertically oriented tubes is created by ramping down the power in the vertically propagating beam pair with about 60 atoms in the center tube. Typical trapping frequencies for the tubes are $\omega_{r,z} = 2 \pi \times (12300(200), 21.9(3))$ Hz along the transversal and longitudinal directions, respectively. In view of our inhomogeneous system we calculate $\gamma$, for a given tube, by assuming a 1D Thomas-Fermi distribution and taking the center density. The reported $\gamma$ is a weighted average over all tubes. We calibrate the depth of the lattice along the tubes by the pulsed Raman-Nath technique\cite{Gould1986}.

\subsection{Commensurability.}
To observe the pinning transition it is not necessary to fulfill the condition of commensurability precisely\cite{Buechler2003}. A finite commensurability parameter $Q=2\pi(n-n_\text{c})$ corresponds to a shift $\delta \mu$ of the chemical potential. Here, $n_\text{c}=2/\lambda$ is the commensurate 1D density. The system stays locked to the MI phase as long as $\delta \mu$ remains smaller than the energy necessary to add another atom. When $Q$ rises beyond a critical value $Q_\text{c}(\gamma,V)$, the system develops finite density excitations, which destroy the long range order of the MI. We find that, for the array of 1D tubes, the commensurability condition in the superfluid regime is fulfilled best when the total atom number is chosen in such a way that the peak density of the center tube is approximately $1.2\ n_\text{c}$.

\end{document}